\documentclass[dvips]{acta}
\usepackage{supertabular,lscape,epsfig}
\usepackage{amssymb}
\usepackage{amsmath}

\SetPages{0}{0}

\SetVol{58}{2008}

\begin{document}

\begin{Titlepage}
\Title{The properties of XO-5b and WASP-82b redetermined using new high-precision transit photometry and global data analyses} 
\Author{A.~M.~S.~Smith}{Nicolaus Copernicus Astronomical Centre, Bartycka 18, 00-716 Warsaw, Poland\\
e-mail:amss@camk.edu.pl}

\Received{Month Day, Year}
\end{Titlepage}

\Abstract{This paper presents new transit photometry from the Isaac Newton Telescope of two transiting exoplanetary systems, XO-5 and WASP-82. In each case the new transit light curve is more precise than any other of that system previously published. The new data are analysed alongside previously-published photometry and radial velocities, resulting in an improved orbital ephemeris and a refined set of system parameters in each case. The observational baseline of XO-5 is extended by very nearly four years, resulting in a determination of the orbital period of XO-5b to a precision of just 50~ms. The mass and radius of XO-5b are $1.19\pm0.03$ and $1.14\pm0.03$ times those of Jupiter, respectively. The light curve of WASP-82 is only the second published for this system. The planetary mass is $1.25\pm0.05 M_{\rm Jup}$, and the radius is $1.71\pm0.08 M_{\rm Jup}$.

}
	     {planetary systems --
   	      planets and satellites: fundamental parameters --
               stars: individual: WASP-82 -- 
               stars: individual: XO-5}

\section{Introduction}

The study of transiting exoplanets (for which radii and absolute masses may be measured) plays a vital role in increasing our understanding of the varied range of extant planets, and of the formation and evolution of planetary systems. Since the discovery of the first transiting exoplanet (Charbonneau et al. 2000; Henry et al. 2000), more than one thousand such detections have been made\footnote{http://www.exoplanet.eu; http://www.astro.keele.ac.uk/jkt/tepcat}, the majority by the {\it Kepler} satellite (Borucki et al. 2010). Although {\it Kepler's} discoveries have revolutionised our understanding of the diverse Galactic planet population, most transiting planets orbiting bright ($V \lesssim 12$) stars were discovered by ground-based surveys, the most prolific of which are the Wide Angle Search for Planets (WASP; Pollacco et al. 2006) and the Hungarian-made Automated Telescope Network (HATNet; Bakos et al. 2002). Bright systems are important because planetary masses and radii can be determined to high precision, and further characterisation, not always possible for fainter stars, is possible. Examples of characterisation observations requiring a relatively bright host star include high-precision radial velocity (RV) measurements to determine orbital obliquities and measurements of planetary transmission and emission spectra to infer atmospheric properties (e.g. Winn 2011).

In this paper, I present new high-precision transit light curves of two bright planetary systems discovered from the ground, and analyse these new data alongside previously published photometry and RVs to refine our knowledge of the system parameters. XO-5 b orbits a slightly evolved G8 star and was discovered by the XO project (Burke et al. 2008), as well as independently by P\'al et al. (2009) using HAT-Net. The planet is similar in both mass and radius to Jupiter, with a density about 80 per cent that of Jupiter, and orbits its host star in a little over four days. Photometry from the two discovery papers were jointly analysed in a third paper (Southworth 2010), and further transit light curves of the system have been published by Maciejewski et al. (2011) and Sada et al. (2012). WASP-82 b is an inflated ($1.7~R_{\rm Jup}$) planet orbiting an evolved F5 star every 2.7~d; its discovery was reported by West et al. (2013). The current work is the first since the discovery paper to report data on this system.

\section{New observations}
\subsection{XO-5}
I observed the transit of XO-5 by its planet, XO-5b, on 2014 October 29/30, using the 2.5-m Isaac Newton Telescope (INT) at the Observatorio del Roque de los Muchachos on La Palma, Spain. The telescope's Wide Field Camera (WFC) was employed in fast readout mode, the exposure time was 60~s for each image, and a Sloan $r^\prime$ filter was used. The telescope was defocused, blurring the stellar point spread function and spreading it over a large ($\sim30$ pixel radius) area of the CCD, so as to reduce the noise associated with flat-fielding errors (e.g. Southworth et al. 2009). Unfortunately, defocusing the INT in this way renders the telescope's autoguider inoperable, but an autoguiding script that updates the telescope pointing based on analysis of the previous science frame was used to keep the position of the target star constant on the CCD to within just a few pixels. The observations cover the entire transit, except for a break between 03:35 and 04:05 UT (phases 1.0047 to 1.0098) due to a failure of the autoguiding script.

The data (from WFC's CCD4, which has a field of view of $22^\prime \times 11^\prime$) were reduced using standard {\sc IRAF} routines to subtract biases, implement a flat-field correction, and perform differential aperture photometry. The photometric aperture radius and selection of reference stars were chosen on the basis of minimising the out-of-transit rms. An aperture radius of 40 pixels (corresponding to $13.2^{\prime\prime}$) was adopted, and five reference stars were used (the fluxes were summed). The reference stars have USNO-B1.0 $R$-band magnitudes of 11.0, 11.0, 11.1, 13.0 and 13.4, and XO-5 has $R$=11.4. The Julian dates provided in the FITS headers were converted to barycentric Julian dates, in the UTC standard.

\subsection{WASP-82}
The INT/WFC was used to observe WASP-82b transiting its host star on 2014 October 30/31, with the same instrumental setup as described in Sec.~2.1, except that a Str\"omgren~$y$ filter was used, and the exposure time was increased to 90~s. Observations cover the transit from around the end of ingress until shortly after the end of egress (phases 0.968 to 1.055). The data were reduced in the same way as described in Sec.~2.1 for XO-5, but the photometric aperture radius was 35 pixels (corresponding to $11.55^{\prime\prime}$) and only two reference stars ($R$=10.7, 10.9, c.f. $R$=9.8 for WASP-82) were used.

\section{Archival data}
Several datasets from previously published papers were included in the analysis, these are described below and the photometry is summarised in Table~1.
\subsection{XO-5}
High-precision light curves from Burke et al. (2008), P\'al et al. (2009), Maciejewski et al. (2011), and Sada et al. (2012) were analysed alongside the new INT light curve. Further, these data were combined with radial velocity measurements from the High-Resolution Spectrograph (HRS) on the Hobby-Eberly Telescope (HET)  from Burke et al. (2008), and from the High Resolution Echelle Spectrometer (HIRES) on Keck-I (P\'al et al. 2009). 

Southworth (2010) chose to omit the lower-quality survey photometry of both Burke et al. (2008) and P\`al et al. (2009) from his analysis. Indeed, Burke et al. (2008) did not use their XO survey photometry, nor their `extended team' light curves (resulting from a collaboration of professional and amateur astronomers) in determining their best-fitting transit model. Instead, they used this data only to determine the orbital ephemeris of the system. Given that there is now sufficient high-precision photometry to constrain well the orbital ephemeris, I too excluded the survey photometry from my analysis. I did, however, opt to include the $i^\prime$ light curve published by P\'al et al. (2009) that covers only about half of the transit, which was omitted by Southworth (2010) for that reason.

\subsection{WASP-82}
The INT light curve presented in this paper was combined with the single high-precision light curve and WASP discovery photometry from West et al. (2013). The WASP photometry was included because there are only two high-precision transit light curves, it therefore plays a significant role in constraining the orbital ephemeris. Also included in the analysis were the radial velocities published by West et al. (2013), from the CORALIE and SOPHIE spectrographs, which are mounted on the 1.2-m Euler Telescope and the 1.93-m Observatoire de Haute Provence (OHP) telescope, respectively.

\MakeTable{cllcccl}{12.5cm}{Summary of the high-precision photometry analysed}
{\hline
ID &Target & Date (UT) & Telescope / Instrument & Filter &RMS$^*$& Ref.$^\dagger$ \\
&&&&&/ppm&\\
\hline
i & XO-5 & 2008 Jan 20 & 1.8-m Perkins / PRISM & $R$ &1250 & B08\\
ii & XO-5 & 2008 Jan 20 & FLWO 1.2-m / KeplerCam & $z^\prime$ &1689& P09\\
iii & XO-5 & 2008 Feb 10 & FLWO 1.2-m / KeplerCam & $i^\prime$ &926& P09\\
iv & XO-5 & 2008 Mar 27 & FLWO 1.2-m / KeplerCam & $i^\prime$ &884& P09\\
v & XO-5 & 2009 Nov 24 & KPNO 2.1-m / FLAMINGOS & $J$ &1596& S12\\
vi & XO-5 & 2010 Jan 21/22 & Calar Alto 2.2-m / CAFOS & $R$ &1107& M11\\
vii & XO-5 & 2010 Nov 11 & Calar Alto 2.2-m / CAFOS & $R$ &688& M11\\
viii & XO-5 & 2014 Oct 30 &  INT/WFC & $r^\prime$ &603& This work\\
ix & WASP-82 & 2012 Nov 21 & 1.2-m SwissEuler / EulerCam & $R$ &767& W13\\
x & WASP-82 & 2014 Oct 31 &  INT/WFC & $y$ &760& This work\\
\hline
\multicolumn{7}{p{11cm}}{$^*$Scatter of binned light curve (bin width = 120 s, as plotted in Figure.~1).\newline
$^\dagger$ B08: Burke et al. (2008); P09: P\'al et al. (2009); S12: Sada et al. (2012); \newline M11: Maciejewski et al. (2011); W13: West et al. (2013).}
}

\section{Data Analysis}

For each system, the available photometry and radial velocities were analysed simultaneously, using a Markov Chain Monte Carlo (MCMC) code. I used the current version of the MCMC code described in Collier Cameron et al. (2007) and Pollacco et al. (2008), which uses the empirical relation between stellar mass, effective temperature, density and metallicity of Southworth (2011). A brief, up-to-date description of the code can be found in Smith et al. (2014). The stellar properties reported in P\'al et al. (2009) were used as a starting point for the MCMC analysis of XO-5, and similarly those reported by West et al. (2013) were used for WASP-82.

Circular orbits were assumed for both XO-5b and WASP-82b, since neither planet has previously been reported as exhibiting any significant orbital eccentricity. Furthermore, there are good reasons, both empirical and theoretical, to expect the orbits of planets such as these to be circular (e.g. Anderson et al. 2012). An additional fit with orbital eccentricity as a free parameter was performed for each system, in order to calculate an upper limit to the eccentricity (Tables 2 \& 3).

One radial velocity datum (from Burke et al. 2008) was excluded from the analysis, since the observation occurred during transit. Similarly, one such point is excluded for WASP-82, as it was by West et al. (2013). P\'al et al. (2009) included an extra term which is dependent upon the bisector span in their fit to their radial velocities; this has the effect of reducing the RV residuals to the fit. Instead, I chose to add `jitter' to the radial velocity uncertainties. Jitter equal to 8 ms$^{-1}$ was added in quadrature to the HIRES radial velocity uncertainties in order for the reduced-$\chi^2$ of that dataset to be approximately unity. No such jitter was required for any other radial velocity dataset.

The MCMC code accounts for stellar limb darkening using a four-coefficient, non-linear model, and the coefficients tabulated by Claret (2000; 2004). The coefficients interpolated for our best-fitting stellar properties are displayed in Table~4 for both XO-5 and WASP-82 in each relevant bandpass. For WASP-82, the WASP photometry was fitted using $R$-band limb-darkening coefficients, as has been done for previous WASP datasets (e.g. Smith et al. 2014).

The best-fitting system parameters describing each system are presented in Tables 2 \& 3, for XO-5 and WASP-82 respectively, and the photometry and radial velocities are shown, along with model fits, in Figure~1. Each light curve was binned in phase with a bin width corresponding to 120~s prior to plotting. The scatter of each binned light curve from the best-fitting model was also calculated, these are tabulated in Table~1. We see that the INT light curve of XO-5 presented in this work has the smallest scatter of all those published on this system, whilst the INT light curve of WASP-82 has a scatter almost identical to that of the only other published high-precision light curve of this system. The scatters of the binned light curves are compared because direct comparison of the rms scatter of light curves with different cadences / bin widths is not meaningful.

As a check, an additional MCMC analysis was performed for each system, in which the only photometry analysed was the new INT light curve. The orbital period was fixed to the value obtained from modelling all of the photometry. In each case, the system parameters resulting from the fit to a single light curve are in excellent agreement with those obtained from the corresponding global analysis, but with increased uncertainties. This is in line with expectations (e.g. Smith et al. 2012).

\begin{figure}[htb]
\begin{minipage}{0.5\textwidth}
\centering
\includegraphics{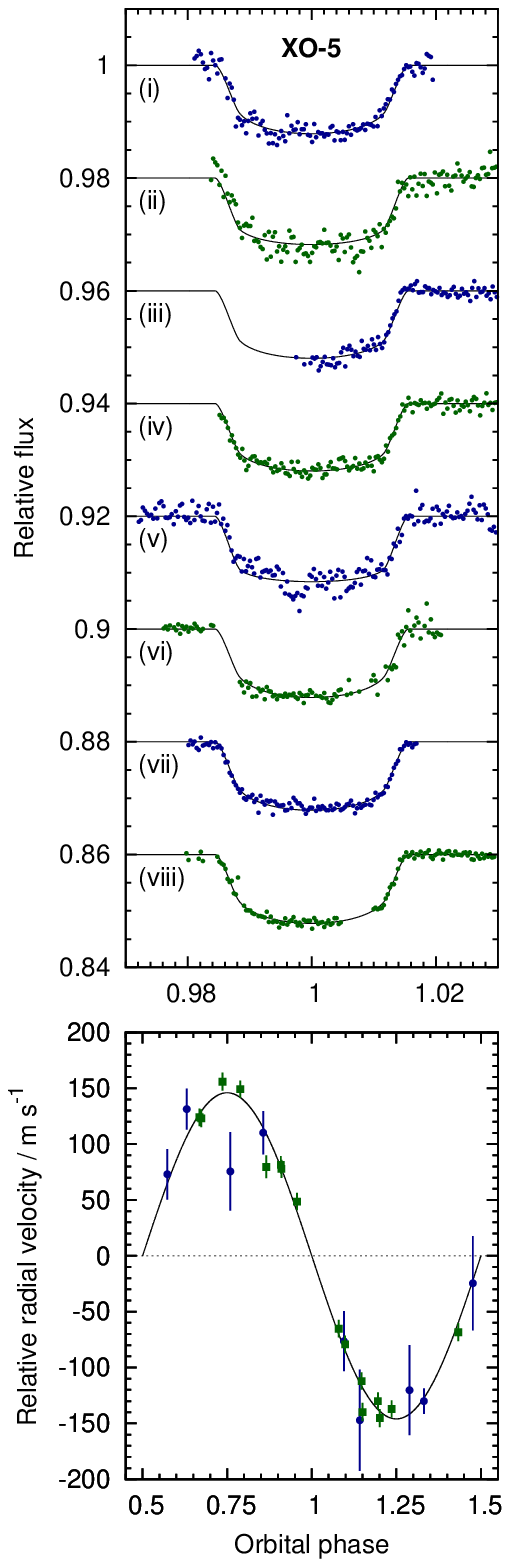}
\end{minipage}
\begin{minipage}{0.5\textwidth}
\centering
\includegraphics{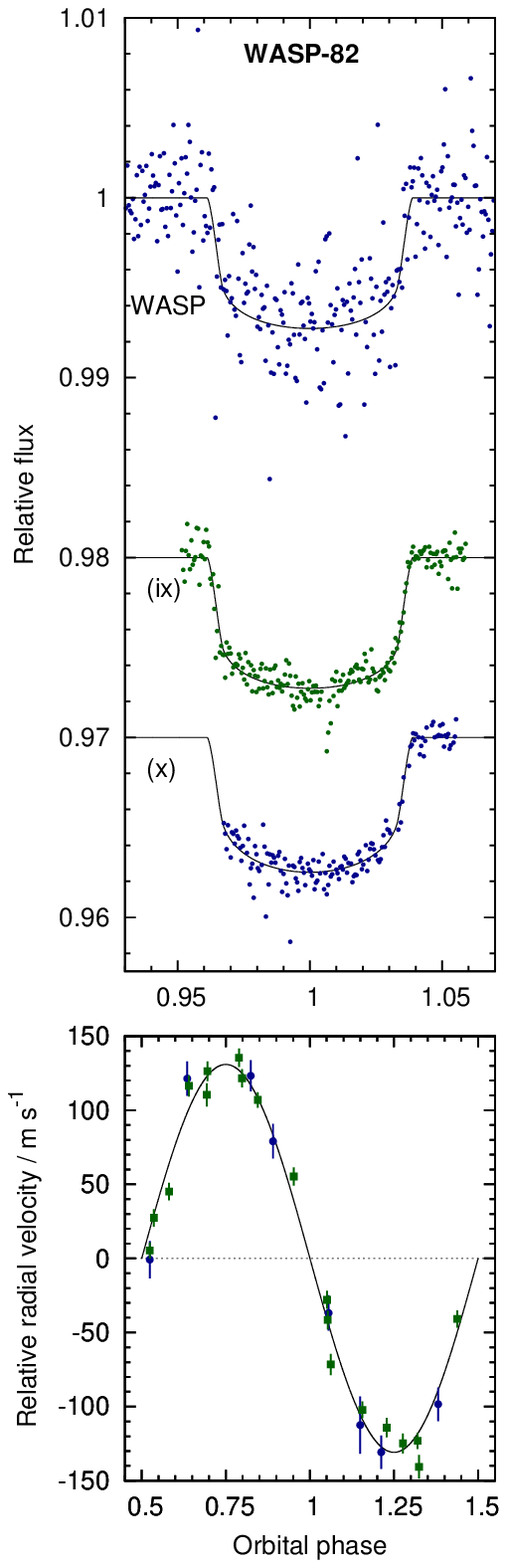}
\end{minipage}
\FigCap{Photometry and radial velocities for XO-5 ({\it left}) and WASP-82 ({\it right}).\newline
{\it Upper panels}: Photometry. Each light curve is offset in flux for clarity, and binned using a 120~s bin width. The light curves are identified using the roman numerals that appear in Table~1. Our best fitting model is shown with a solid line.\newline
{\it Lower panel}: Radial velocities. HET/HRS and OHP/SOPHIE data are displayed as blue circles; Keck/HIRES and Euler/CORALIE data are represented by green squares. The jitter added to the HIRES uncertainties is not shown. The centre-of-mass velocities have been subtracted, as have the fitted offsets between the datasets.}
\end{figure}

\begin{landscape}
\MakeTable{lccccc}{12.5cm}{System parameters: XO-5}
{\hline
Parameter & Symbol & Unit & This work & Literature & Ref.$^\dagger$\\
\hline 
Orbital period	    	    	    	    & 	$P$ & d &				$4.1877558\pm0.0000006$ & $4.187754\pm0.000002$ & M11 \\
Epoch of mid-transit	    	    	    & 	$T_{\rm c}$ &BJD, UTC &			$2456864.3129\pm0.0002$ & $2454485.6677\pm0.0003$ & M11 \\
Transit duration    	    	    	    & 	$T_{\rm 14}$ &d &			$0.1299\pm$0.0007 & $0.131\pm0.001$ & P09 \\
Planet-to-star area ratio   	    	    & 	$\Delta F=R_{\rm P}^{2}/R_{*}^{2}$&...&$0.0108\pm0.0001$ & ...& S10 \\
Planet-to-star radius ratio   	    	    & 	$R_{\rm P}/R_{*}$&...&$0.1039\pm0.0007$ & $0.106\pm0.003$& B08 \\
&&&& $0.1050\pm0.0009$& P09 \\
&&&& $0.101\pm0.002$& M11 \\
&&&& $0.102\pm0.003$& S12 \\
Transit impact parameter    	    & 	$b$ &...&					$0.55\pm0.03$ & $0.56^{+0.03}_{-0.05}$ & P09 \\
Stellar orbital velocity semi-amplitude     & 	$K_{\rm *}$ &m s$^{-1}$ &		$146\pm3$ & $145\pm2$ & P09 \\
Stellar effective temperature         &  $T_{\rm *, eff}$ & K &				$5430\pm70$ & $5370\pm70$ & P09 \\
Stellar metallicity			    &  [Fe/H] & dex &				$+0.05\pm0.06$ & $+0.05\pm0.06$ & P09 \\

Ingress / egress duration    	    & 	$T_{\rm 12}=T_{\rm 34}$ &d &			$0.0170\pm0.0008$ & $0.018\pm0.001$ & P09 \\
Orbital inclination angle   	    	    & 	$i_P$ &$^\circ$  &			$86.8\pm0.2$ & $87.0\pm0.7$ & S10 \\
Orbital eccentricity (adopted)	    	    	    & 	$e$ &...&	 		$0  $ & $ 0 $ & P09 \\
Orbital eccentricity (3-$\sigma$ upper-limit) &... &...&				$0.06$ & $...$ & ... \\
Stellar mass	    	    	    	    & 	$M_{\rm *}$ & $M_{\rm \odot}$ &		$1.04\pm0.03$ & $0.91\pm0.06$ & S10 \\
Stellar radius	    	    	    	    & 	$R_{\rm *}$ & $R_{\rm \odot}$ &		$1.13\pm0.03$ & $1.07\pm0.06$ & S10 \\
log (stellar surface gravity)     	    & 	$\log g_{*}$ & (cgs) &			$4.35\pm0.02$ & $4.34\pm0.04$ & S10 \\
Stellar density     	    	    	    & 	$\rho_{\rm *}$ &$\rho_{\rm \odot}$ &	$0.72\pm0.04$ & $0.8\pm0.1$ & S10 \\
Planet mass 	    	    	    	    & 	$M_{\rm P}$ &$M_{\rm Jup}$ &		$1.19\pm0.03$ & $1.08\pm0.06$ & S10 \\
Planet radius	    	    	    	    & 	$R_{\rm P}$ &$R_{\rm Jup}$ &		$1.14\pm0.03$ & $1.09\pm0.08$ & S10 \\
Planet density	    	    	    	    & 	$\rho_{\rm P}$ &$\rho_{\rm J}$ &	$0.80\pm0.06$ & $0.8\pm0.2$ & S10 \\
Orbital major semi-axis     	    	    & 	$a$ &au  &				$0.0515\pm0.0005$ & $0.049\pm0.001$ & S10 \\
Planet equilibrium temperature$^\ddag$ & 	$T_{\rm P, A=0}$ &K &				$1230\pm20$ & $1220\pm30$  & P09\\
\hline
\multicolumn{6}{p{17cm}}{$^\dagger$ B08: Burke et al. (2008); P09: P\'al et al. (2009); S10: Southworth (2010);  M11: Maciejewski et al. (2011); S12: Sada et al. (2012); W13: West et al. (2013). $^\ddag$ Assuming uniform heat redistribution}
}
\end{landscape}

\begin{landscape}
\MakeTable{lcccc}{12.5cm}{System parameters: WASP-82}
{\hline
Parameter & Symbol & Unit & This work & West et al. (2013) \\
\hline 
Orbital period	    	    	    	    & 	$P$ & d &				$2.705785\pm0.000002$ & $2.705782\pm0.000003$ \\
Epoch of mid-transit	    	    	    & 	$T_{\rm c}$ &HJD, UTC &			$2456931.8444\pm0.0008$ & $2456157.9898\pm0.0005$ \\
Transit duration    	    	    	    & 	$T_{\rm 14}$ &d &			$0.208\pm0.001$ & $0.208\pm0.001$ \\
Planet-to-star area ratio   	    	    & 	$\Delta F=R_{\rm P}^{2}$/R$_{*}^{2}$&...&$0.0063\pm0.0001$ & $0.0062\pm0.0001$ \\
Transit impact parameter    	    & 	$b$ &...&					$0.3^{+0.1}_{-0.2}$ & $0.2\pm0.1$ \\
Stellar orbital velocity semi-amplitude     & 	$K_{\rm *}$ &m s$^{-1}$ &		$131\pm2$ & $131\pm2$ \\
System velocity     	    	    	    &     	$\gamma$ &km s$^{-1}$ &		$-23.62828\pm0.00006$ & $-23.62827\pm0.00007$ \\
Stellar effective temperature         &  $T_{\rm *, eff}$ & K &				$6480\pm90$ & $6490\pm100$ \\
Stellar metallicity			    &  [Fe/H] & dex &				$+0.1\pm0.1$ & $+0.1\pm0.1$ \\

Ingress / egress duration    	    & 	$T_{\rm 12}=T_{\rm 34}$ &d &			$0.016\pm0.001$ & $0.0156^{+0.0012}_{-0.0004}$ \\
Orbital inclination angle   	    	    & 	$i_P$ &$^\circ$  &			$86.6\pm 2$ & $88^{+1}_{-2}$ \\
Orbital eccentricity (adopted)	    	    	    & 	$e$ &...&	 		0 & 0 \\
Orbital eccentricity (3-$\sigma$ upper-limit) &... &...&				0.05 & $0.06$ \\
Stellar mass	    	    	    	    & 	$M_{\rm *}$ & $M_{\rm \odot}$ &		$1.64\pm0.08$ & $1.63\pm0.08$ \\
Stellar radius	    	    	    	    & 	$R_{\rm *}$ & $R_{\rm \odot}$ &		$2.219^{+0.1}_{-0.08}$ & $2.18^{+0.08}_{-0.05}$ \\
log (stellar surface gravity)     	    & 	$\log g_{*}$ & (cgs) &			$3.96\pm0.03$ & $3.97^{+0.01}_{-0.02}$ \\
Stellar density     	    	    	    & 	$\rho_{\rm *}$ &$\rho_{\rm \odot}$ &	$0.15\pm0.02$ & $0.158^{+0.006}_{-0.014}$ \\
Planet mass 	    	    	    	    & 	$M_{\rm P}$ &$M_{\rm Jup}$ &		$1.25\pm0.05$ & $1.24\pm0.04$ \\
Planet radius	    	    	    	    & 	$R_{\rm P}$ &$R_{\rm Jup}$ &		$1.71\pm^{+0.09}_{-0.07}$ & $1.67^{+0.07}_{-0.05}$ \\
log (planet surface gravity)     	    & 	$\log g_{\rm P}$ & (cgs) &		$2.99^{+0.03}_{-0.04}$ & $3.01^{+0.02}_{-0.03}$ \\
Planet density	    	    	    	    & 	$\rho_{\rm P}$ &$\rho_{\rm J}$ &	$0.25\pm0.03$ & $0.27^{+0.02}_{-0.03}$ \\
Orbital major semi-axis     	    	    & 	$a$ &au  &				$0.0448\pm0.0007$ & $0.0447\pm0.0007$ \\
Planet equilibrium temperature    & 	$T_{\rm P, A=0}$ &K &				$2200\pm50$ & $2190\pm40$ \\
~~~~~(uniform heat redistribution)&&&&\\
\hline
}
\end{landscape}

\MakeTable{lccrrrr}{12.5cm}{Stellar limb-darkening coefficients}
{\hline
Star 		& Light curves$^\dag$ 		& Band & $a_1$~ &  $a_2$~  &  $a_3$~ & $a_4$~  \\
\hline 
XO-5 & i, vi, vii & $R$ & 0.702 & -0.561 & 1.210 & -0.561\\
XO-5 & ii & $z^\prime$ & 0.778 & -0.750 & 1.145 & -0.508\\
XO-5 & iii, iv & $i^\prime$ & 0.766 & -0.716 & 1.235 & -0.557\\
XO-5 & v & $J$ & 0.601 & -0.131 & 0.279 & -0.166\\
XO-5 & viii & $r^\prime$ & 0.678 & -0.504 & 1.188 & -0.561\\
WASP-82 & ix & $R$ & 0.483 & 0.479 & -0.350 & 0.082\\
WASP-82 & x & $y$ & 0.415 & 0.685 & -0.459 & 0.123\\
\hline
\multicolumn{7}{p{7cm}}{$^\dag$Light curve IDs correspond to those listed in Table~1.}
}

\section{Discussion and Conclusions}

This work represents the most comprehensive data analyses yet performed on both the XO-5 and WASP-82 planetary systems. Tables 2 and 3 present the system parameters derived in this work alongside previously-published values for comparison. For WASP-82, these literature values all come from West et al. (2013), whereas for XO-5 they come from several different papers. Most parameters are from Southworth (2010), who analysed photometry from the two independent system discovery papers, however some parameters are not reported in that work, so we turn to P\'al et al. (2009) for these values. Finally, the best and most recent orbital ephemeris prior to the current work was published by Maciejewski et al. (2011), so their values of $P$ and $T_{\rm c}$ are used as a comparison for those determined herein.

In the case of XO-5, most parameters are in close ($< 1 \sigma$) agreement with previously published values, but have smaller uncertainties. By extending the baseline of photometric observations of the system by very nearly four years, the orbital ephemeris is significantly improved; the uncertainty on the orbital period is now just 50~ms. 

The ratio of planetary to stellar radius, derived from observations, is in excellent agreement with all previous work (four values of this quantity are listed in Table~2 -- one from each previous XO-5 observational paper). The planetary and stellar masses found in this work, however, are each slightly larger than those found by Southworth (2010); they are discrepant by around 1.7~$\sigma$. This results in a slightly larger stellar radius too, discrepant at 1.0~$\sigma$, although the stellar parameters I report are closest to those of Burke et al. (2008). This reflects the fact that very often the precision to which fundamental planetary properties can be determined is limited by our knowledge of the host star's properties (e.g. Torres et al. 2009). In conclusion, this paper presents a self-consistent determination of the parameters, from an analysis of all available RVs and high-precision photometry, something that has not previously been done for this system since the first announcement of this planet (Burke et al. 2008).

In the case of WASP-82, the parameters from the new analysis are in very close agreement with those of West et al. (2013). The uncertainty on the orbital period is somewhat reduced, and those on other parameters are very similar to those previously published. The new light curve has confirmed the veracity of the analysis reported in West et al. (2013), which was only able to draw upon a single high-precision transit light curve.

Good knowledge of system parameters is vital both for observers planning future characterisation observations, and for theorists seeking to unmask patterns and correlations amongst the ever-increasing ensemble of known exoplanetary systems. In particular, parameters based on a complete analysis of all available data are preferable in this respect to several competing, disparate investigations. In this paper, comprehensive analyses of XO-5 and WASP-82 were presented.

\Acknow{The Isaac Newton Telescope is operated on the island of La Palma by the Isaac Newton Group (ING) in the Spanish Observatorio del Roque de los Muchachos of the Instituto de Astrof\'{i}sica de Canarias (IAC); observations were made under programme I/2014B/12. The author wishes to thank the staff of the IAC and the ING for their assistance, particularly Teo Mo\v{c}nik for his help with the observations.  The author also wishes to extend his gratitude to Andrew Collier Cameron for permitting the use of his MCMC code, and to Ernst de Mooij and James McCormac for making their autoguiding script available for use during the observations. This work was supported by the Polish NCN through grant no. 2012/07/B/ST9/04422.
}


\begin{references}

\refitem{Anderson, D. R., et al.}{2012}{\MNRAS}{422}{1988}
\refitem{Bakos, G. \'A., L\'az\'ar, J., Papp, I., S\'ari, P., \& Green, E. M.}{2002}{\ApJ}{669}{1167}
\refitem{Borucki, W. J., et al.}{2010}{Science}{327}{977}
\refitem{Burke, C. J., et al.}{2008}{\ApJ}{686}{1331}
\refitem{Charbonneau, D., Brown, T. M., Latham, D. W., \& Mayor, M.}{2000}{\ApJ}{529}{L45}
\refitem{Claret, A.}{2000}{A\&A}{363}{1081}
\refitem{Claret, A.}{2004}{A\&A}{428}{1001}
\refitem{Collier Cameron, A., et al.}{2007}{\MNRAS}{380}{1230}
\refitem{Henry, G. W., Marcy, G. W., Butler, R. P., \& Vogt, S. S.}{2000}{\ApJ}{529}{L41}
\refitem{Maciejewski, G., Seeliger, M., Adam, C., Raetz, S., \& Neuh\"auser, R.}{2011}{\Acta}{61}{25}
\refitem{P\'al, A., et al.}{2009}{\ApJ}{700}{783}
\refitem{Pollacco, D., et al.}{2006}{\PASP}{118}{1407}
\refitem{Pollacco, D., et al.}{2008}{\MNRAS}{385}{1576}
\refitem{Sada, P. V., et al.}{2012}{\PASP}{124}{212}
\refitem{Smith, A. M. S., et al.}{2012}{\AJ}{143}{81}
\refitem{Smith, A. M. S., et al.}{2014}{A\&A}{570}{A64}
\refitem{Southworth, J., et al.}{2009}{\MNRAS}{396}{1023}
\refitem{Southworth, J.}{2010}{\MNRAS}{408}{1689}
\refitem{Southworth, J.}{2011}{\MNRAS}{417}{2166}
\refitem{Torres, G., Winn, J., \& Holman, M. J.}{2009}{IAUS}{253}{482}
\refitem{West, R. G., et al.}{2013}{arXiv:1310.5607}{}{}
\refitem{Winn, J. N.}{2011}{Exoplanet Transits and Occultations, ed. S. Seager}{55}{}

\end{references}
\end{document}